\documentclass[prl,showpacs,twocolumn]{revtex4}

\usepackage{hyperref,fontenc,times,amsmath,amsfonts,amssymb,graphics,graphicx,epsfig,color,bbm,natbib,bm}

\newcommand{\dd}{\mathrm{d}}
\newcommand{\ee}{\mathrm{e}}
\newcommand{\ii}{\mathrm{i}}
\newcommand{\Ai}{\ensuremath{\operatorname{Ai}}}

\begin{document}

\title{Scaling laws for precision in quantum interferometry and bifurcation landscape of optimal state}
\author{Sergey Knysh}\email{Sergey.I.Knysh@nasa.gov}

\author{Vadim N. Smelyanskiy}\email{Vadim.N.Smelyanskiy@nasa.gov}

\author{Gabriel A.\ Durkin} \email{gabriel.durkin@qubit.org}
\affiliation{Quantum Laboratory, Applied Physics Center, NASA Ames Research Center,  Moffett Field, California 94035, USA}
\date{\today}
\begin{abstract}
Phase precision in optimal 2-channel quantum interferometry is studied in the
limit of large photon number $N\gg 1$, for  losses occurring in
either one or both channels. For losses in one channel an optimal state
undergoes an intriguing sequence of local bifurcations as the losses
or the number of photons increase. We further show that fixing the loss paramater
determines a scale for quantum metrology -- a crossover  value of the photon number 
$N_c$ beyond which the supra-classical precision is progressively lost. 
For large losses the optimal state also has a different structure from those considered previously.
\end{abstract}
\pacs{42.50.-p,42.50.St,06.20.Dk}

\maketitle

It has been recognized that using quantum states of light may increase the resolution of interferometric measurements
\cite{Kok02,Giovanetti04,Nagata07}. Particular states of $N$ photons
achieve the Heisenberg limit of phase resolution for standard error on
the phase estimate $\Delta \varphi = 1 / N$, an improvement over the
classical (or shot-noise) limit  $\Delta \varphi = 1 / \sqrt{N}$  that is obtainable when $N$ photons enter the interferometer one at a time.  These bounds are derived by an application of
the Cramer-Rao inequality \cite{Giovanetti04} for the standard error
of an unbiased estimator, $\Delta \varphi \geq (\nu
\mathcal{F})^{-1/2}$, where $\mathcal{F}$ is the quantum Fisher
information (QFI) \cite{Braunstein94} and $\nu $ is the number of
repeated independent trials. Assuming any instrument is composed of
three components: quantum input state, dynamics and
measurement; the functional $\mathcal{F}$  depends only on the first
two --- it assumes an optimal measurement choice. For pure states in a
single mode $\mathcal{F}/4 = \Delta^2 \hat{n} \equiv \langle \hat n^2
\rangle-\langle \hat n\rangle^2$ (where $\hat{n}$ is the number
operator) and a familiar uncertainty relation is recovered: $\Delta n
 \Delta \varphi \geq 1/2$.  Thus, for a lossless two-mode interferometer QFI and precision are greatest for the maximum variance state, or `NOON state'; it saturates the Heisenberg limit. Unfortunately, it is also highly susceptible to noise, especially dissipation \cite{Chen07}.

To mitigate this problem various two-component states were proposed  \cite{Huver08,DemkowiczDobrzanski09,Dorner09},  where the  loss
of a number of photons in the first mode  does not destroy the superposition.
The precision performance under dissipation of various Gaussian states, e.g. squeezed,
coherent and thermal states, has also been considered recently  \cite{gauss}.
In all cases, the precision was found to be supra-classical for certain range of losses and  $N$.

In the lossy case the  pure input state of two oscillator modes maximizing QFI
\begin{equation}
 | \phi \rangle = \sum_{n = 0}^N \phi_n |n\rangle_1|N - n \rangle_2,
\label{eq:input-state}
\end{equation}
must balance supra-classical precision against robustness to photon
loss. In this notation the NOON state has two non-zero components, 
$ \phi_0= \phi_N = 1/\sqrt{2}$. For a lossy
interferometer light propagates in each arm as a damped
harmonic oscillator, with frequencies $\omega^{(1)}, \omega^{(2)}$ and
dissipation $\gamma^{(1)}, \gamma^{(2)}$. Equivalently, losses can be
introduced by beam-splitters in each mode with reflectivity 
$R^{(1,2)}=1 - \exp \{- \gamma^{(1,2)} t \}$. 
Those lost photons siphoned out of the modes are then traced over.
In the simpler case of losses in only one of the two modes, $R^{(1)}=R>0$, $R^{(2)}=0$, as might occur when that mode is directed
through a partially transparent test sample, the state $| \phi \rangle$
decays into a mixture
$\hat\rho = \sum_k | \psi_k \rangle \langle \psi_k |$ with
\begin{equation}
  | \psi_k \rangle = \frac{1}{\sqrt{w_k}} \sum_n \sqrt{\Lambda_{n ; k}}
  \ee^{\ii n \varphi} \phi_n |n - k, N - n \rangle,
\label{psik}
\end{equation}
corresponding to the loss of $k$ photons. Here $w_k$ is the normalization 
factor; the phase differense is $\varphi = \left(\omega^{(1)}-\omega^{(2)}\right)t$
and the loss enters via coefficients $\Lambda_{n;k}=\binom{n}{k} R^k (1-R)^{n-k}$.
Fisher information of the mixed state resulting from losses is a
weighted sum over pure components $\mathcal{F} = \sum_k w_k \mathcal{F}_k$,
where $\mathcal{F}_k = 4 \Delta^2 \hat n_1$ for pure states $|\psi_k\rangle$ \cite{Dorner09}. 

Refs.~\cite{Dorner09,DemkowiczDobrzanski09} use numerical optimization
to construct states $| \phi_\mathrm{opt} \rangle$  maximizing this information
for a range of photon numbers $N$ and loss  $R$. For larger
photon numbers/loss, a complicated multicomponent structure
arises, but the evolution of the optimal state with increasing loss was previously not well-understood.
More importantly, the question of the asymptotic scaling with large $N$ of the Fisher
information has been left unanswered. The significance of this issue becomes apparent 
when one examines the suitability of quantum-enhanced sensors for tasks such as gravitational 
wave observation where $N \gg 1$ is necessary to reach the desired sensitivity. Later we will discuss 
the case of free-space target acquisition and ranging (e.g. quantum LIDAR) where combined loss $R$ 
due to atmospheric attenuation and limited target reflectance is typically over $99\%$.

\begin{figure}[!ht]
\includegraphics[width=3in]{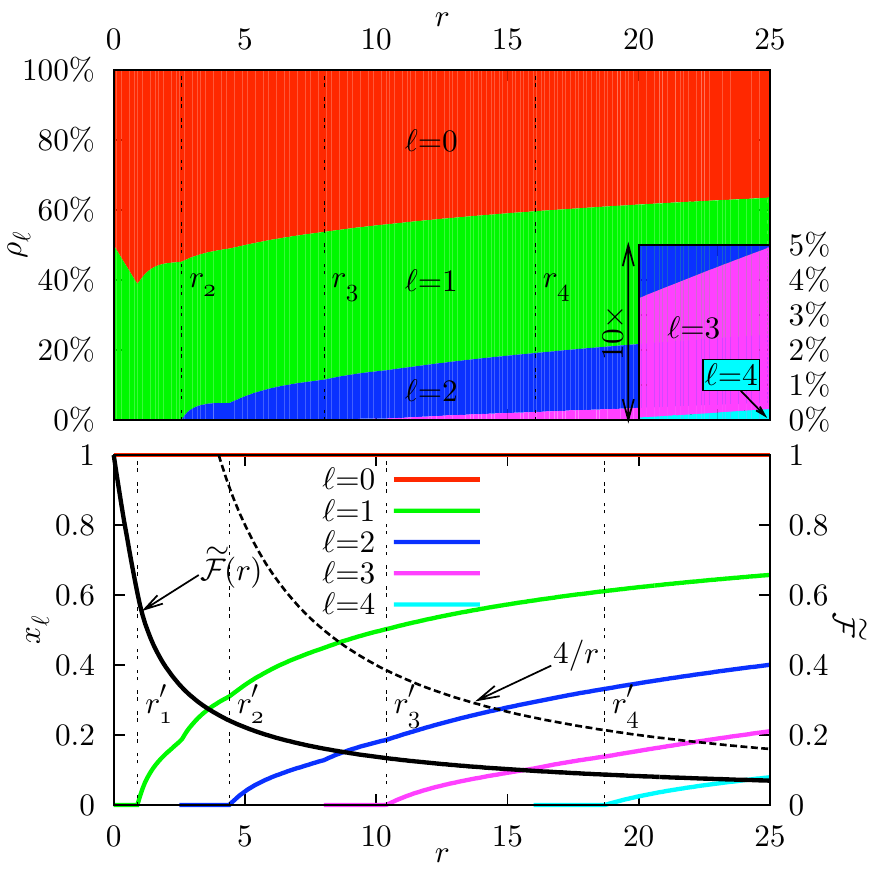}
\caption{\label{fig:pw30}
Probability weights $\rho_\ell$, represented as stacked
histograms (top) and locations $x_\ell$ (bottom), as a function of $r$.
Different components are indicated using color.
Black solid line on the bottom figure is the rescaled Fisher information
$\tilde{\mathcal{F}}(r)$ and the blacked dashed is its asymptote $4/r$;
convergence takes place for much larger values of $r$.
Thresholds $r_2, r_3, r_4$ correspond to appearances of new
components at the origin. Components separate from the origin
at critical values $r'_1, r'_2,r'_3,r'_4$.
The data for $20 \leqslant r \leqslant 25$ is magnified (top figure, lower
right corner) showing components with very small weight.
}
\end{figure}

Here we study the analytically tractable limit $N \gg 1$ by treating $n/N \equiv x \in [0;1]$ as a continuous parameter. Examining the limits of small and large loss has revealed a scaling relationship for the optimal Fisher information
\begin{equation}
  \mathcal{F}_{\textrm{opt}} (N,R) = N^2 \tilde{\mathcal{F}} \left( \frac{NR}{1-R} \right),
\label{scaling}
\end{equation}
that cleanly interpolates between these limits. The non-trivial 
dependence on $N$ and $R$ is
captured by a single quantity: $r = N R / (1-R)$. The structure of
the optimal state also depends on $r$ alone, save for small
differences due to discrete nature of parameter $x=n/N$.
We were able to demonstrate that for any finite $r$ the optimal state 
can have only a finite number of components.
This number increases with $r$ as the optimal state undergoes
a sequence of bifurcations: unbalanced NOON state ceases to be optimal for
$r < r'_1 \approx 0.912957$ \cite{footnote1}, superseded by a state
$\sqrt{1-\rho_1} |N\rangle_1|0\rangle_2 + \sqrt{\rho_1} |x_1 N\rangle_1|(1-x_1)N \rangle_2$
as has been noted in \cite{Dorner09} (a similar state has been proposed in Ref.\cite{Huver08}).
For larger values of the paramater ($r>r_2$), an optimal state acquires a third component 
$|0\rangle_1|N\rangle_2$, which shifts away from the origin to $|x_2 N\rangle_1|(1-x_2)N\rangle_2$
for $r>r'_2$. The universal set of bifurcation points $r'_1<r_2<r_2'<r_3<r'_3<\cdots$ as well
the weights $\rho_\ell$ and locations $x_\ell$ of components in $m+1$-component state
$|\phi\rangle = \sum_\ell \sqrt{\rho_\ell} |x_\ell N\rangle$ are determined by solving a system 
of $2m-1$ or $2m$ equations. The results are shown in Fig.~\ref{fig:pw30}. 
An important caveat is that since $M_\ell = x_\ell N$ are not integers in general, single non-integer 
components split into two adjacent integer components for finite $N$.

The numerical results of Ref.~\cite{Dorner09} correspond to the leftmost region $r \sim 1$ of Fig.~\ref{fig:pw30}. 
In this work we are primarily interested in the regime $r \gg 1$ (not shown).
Our motivation is that the loss parameter $R$ is determined by the properties of the 
medium and is, therefore, fixed. With $R=0$ increasing  $N$
provides continuous improvement in phase precision \emph{per photon} -- we are interested in whether this remains 
the case in the presence of losses. (It should be noted that no efficient procedure for generating optimal 
$N$-photon states for very large $N$ presently exists.)  Many applications, such as 
gravitational wave detectors, pursue orders of magnitude improvement over current precision limits, hence the importance 
of finding out if any fundamental limits to the strategy of increasing the photon number $N$ exist.

With increasing $N$, optimal states seemingly increase in complexity as the value of $r$
and the number of components increases. However, as the density of these components increases, 
for large $N$ the optimal state $|\phi\rangle$ may be approximated by a continuous function.
Examination of Eq.~(\ref{scaling}) shows that, $r$ being proprtional to $N$, a quadratic (Heisenberg) 
scaling of the Fisher information is obtained only for small $r$,
where $\alpha(r)=-\dd(\ln \tilde{\mathcal{F}}) / \dd(\ln r)$ is small, see Fig.~\ref{fig:pw30}.
(Here we are fitting to a local exponential model: $\tilde{\mathcal{F}} \sim r^{-\alpha}$.)
In the opposite limit $r \gg 1$ one has $\tilde{\mathcal{F}} \sim 4/r$, i.e. $\alpha=1$.
so that for $N \gg 1$ the Fisher information increases only linearly, $\mathcal{F} \propto N$. 
Precision, quantified by $\mathcal{F}$, will always improve \emph{at  least} linearly with the 
photon resource $N$, (by sending the photons through the instrument one at a time) but the 
more insightful question is how does the amount of `intrinsic' Fisher information, i.e. \emph{per photon},
scale with $N$ if photons are combined in some optimal quantum
superposition -- indeed,  how does this scaling change, given a loss parameter $R$?

\emph{General Upper Bound:}
This general linear upper bound can be demonstrated without making approximation
of large $N$ or $R \sim 1$. Since the variance is unaffected by a constant shift,
one can rewrite the $\mathcal{F}$ as
\begin{equation}
4 \sum_k \left[ \sum_n \Lambda_{n ; k}  \Big( n - \frac{k}{R}
  \Big)^2 | \phi_n |^2  - 4 w_k  \Big\langle \psi_k \Big| n - \frac{k}{R}
  \Big| \psi_k \Big\rangle^2 \right].
\label{Feq}
\end{equation}
Observing that the second term is negative and performing the sum
over $k$ in the first term, we obtain the inequality
\begin{equation}
  \mathcal{F} \leqslant 4 \frac{1 - R}{R} \sum_n n| \phi_n |^2 \leqslant 4
  \frac{1 - R}{R} N = \mathcal{F}_{\textrm{upper}}. \label{Fmax}
\end{equation}
This upper bound ($\mathcal{F}_{\textrm{upper}}$) is always valid for any $R>0$ demonstrating that 
quadratic precision (at the Heisenberg limit $\mathcal{F} \propto N^2$) is only possible for $R \propto 1/N$. 
When $R$ is fixed, it implies $\mathcal{F} \propto N$, scaling proportional to the shot-noise limit. 
This bound also appeared recently in the complementary context of global phase estimation \cite{KDD}. 

Limit $\mathcal{F}_{\textrm{upper}}$ is reachable asymptotically 
as can be shown by constructing a wavefunction that minimizes the
correction $\Delta \mathcal{F}=\mathcal{F}_{\textrm{upper}} -\mathcal{F}$.
In the limit $R \sim 1$ we approximate the true optimal state $\{\phi_n \}$
by a continuous function $\tilde{\phi} (\tilde{x})$ (with $\tilde{x} =
1-x = 1 - n / N$), smooth on scales $\sim 1 / \sqrt{N}$ and obtain approximately
\begin{equation}
  \Delta \mathcal{F} \approx \frac{4 N^2}{r} \int_0^{\infty} \left[
  \tilde x \tilde{\phi}^2 (\tilde x) + \frac{4}{r}  
  \tilde{\phi}'^2 (\tilde x) \right] \dd \tilde x,
  \label{DF}
\end{equation}
where the upper limit has been set to infinity since the width of
$\tilde\phi(\tilde x)$ is much smaller than $1$. The term proportional
to $\tilde\phi^2(\tilde x)$ is the first term of Eq.~(\ref{Feq}) subtracted from
$\mathcal{F}_{\textrm{upper}}$, and the term proportional to
$\tilde{\phi}'^2(\tilde x)$ is the second term in Eq.~(\ref{Feq})
taken with the opposite sign.

Minimization of (\ref{DF}) subject to the boundary condition
$\phi (0) = 0$ \cite{footnote2}
and the normalization constraint produces
\begin{equation}
  \tilde{\phi} (\tilde x) = \frac{(r / 4)^{1 / 6}}{\Ai' (\mu_1)} \Ai
  \left( \left( \frac{r}{4} \right)^{1 / 3} \tilde x + \mu_1 \right),
\label{airy}
\end{equation}
where $\Ai (z)$ is the Airy function, $\mu_1 \approx - 2.338107 \ldots$
is its first (largest) zero, and the prefactor ensures normalization. Together
with the next order correction, the Fisher information of the optimal state
$\mathcal{F}_{\textrm{opt}}=\mathcal{F}_{\textrm{upper}} - \Delta \mathcal{F}_{\min}$ is
\begin{equation}
  \mathcal{F}_{\textrm{opt}} = \frac{4N^2}{r} \left[ 1 - | \mu_1 | \left( \frac{4}{r}
  \right)^{1 / 3} + O\left( \frac{1}{r^{1/2}} \right) \right] .
\label{Fone}
\end{equation}
For $R \sim 1$ ($r \sim N$) the width of (\ref{airy}) is $O(N^{2/3})$;
so is the leading correction in Eq.~(\ref{Fone}). The upper bound
becomes saturated when the number of photons exceeds a value of
$N_c$ estimated by equating the principal term and the leading order
correction in Eq.~(\ref{Fone}). This yields $N_c = r_c (1-R)/R$ with 
$r_c \sim 4 |\mu_1|^3 \sim 50$. 

\emph{Arbitrary Loss in Both Arms:} Whenever both
$R^{(1)}$ and $R^{(2)}$ are non-zero, the 
density matrix is a mixture of pure states $| \psi_{k_1 k_2} \rangle$ resulting 
from the loss of $k_1$ and $k_2$ photons in modes $(1)$ and $(2)$ respectively. 
In Eq.~\eqref{psik} the factor $\sqrt{\Lambda_{n; k}}$  becomes 
$\sqrt{\Lambda_{n ;k_1}^{(1)} \Lambda_{n ; k_2}^{(2)}}$ and the states$| \psi_{k} \rangle$ 
become $| \psi_{k_1 k_2} \rangle$. 

The number of photons lost in each mode is not observed directly,  although their sum $k=k_1+k_2$ can be inferred by subtracting the detected photon number from the input $N$. Consequently the linear decomposition of the Fisher information serves only as
an upper bound $\mathcal{F} \leqslant \sum_{k_1,k_2} w_{k_1k_2} \mathcal{F}_{k_1k_2}$ \cite{DemkowiczDobrzanski09}
and the determination of quantum Fisher information requires the
diagonalization of the density matrix \cite{Braunstein94}:
\begin{equation}
  \mathcal{F} = 4 \sum_i \lambda_i \langle v_i| \hat n_1^2 | v_i \rangle
  - \sum_{\substack{i,j\\\lambda_i, \lambda_j>0}} \frac{8\lambda_i\lambda_j}{\lambda_i+\lambda_j} 
  |\langle v_i| \hat n_1 |v_j \rangle|^2,
\label{Fdiag}
\end{equation}
where $\lambda_i$ and $|v_i\rangle$ are eigenvalues and eigenvectors of the density matrix respectively.
Diagonalizations within subspaces corresponding to a fixed \emph{total} number of lost photons
$k=k_1+k_2$ may be carried out independently. 
In the limit $N \gg 1$ the coefficients $\Lambda_{n;k}^{(1,2)}$ may be approximated by Gaussians so that 
the corresponding density matrix is also Gaussian in the continuous limit as long as $R^{(1)} \neq R^{(2)}$
and the wavefunction $|\phi\rangle$ is smooth on scales $\sim \sqrt{N}$. This density matrix
may be expanded in terms wavefunctions of harmonic oscillator with the aid of Mehler formula \cite{footnote3} 
and the sum (\ref{Fdiag}) is evaluated noting that non-zero matrix elements correspond to $j = i \pm 1$.
The surprising outcome is that the exact Fisher information equals the linear upper bound ($\mathcal{F}_{\textrm{upper}}$ for arbitrary loss in both arms) in the asymptotic limit.
This is also true in the symmetric loss case ($R^{(1)}=R^{(2)}$) as the optimal state itself turns out to be a Gaussian.
This case has some import; firstly, it is relevant for balanced instruments 
where phases may be introduced in either arm, e.g. gyroscopes, and secondly; the analysis has an extended 
applicability beyond losses in modes (1) and (2) to those occurring in any superposition of these modes. 
Accordingly, the discussion is applicable to losses in detection after the mode mixing.

Expressed in terms of parameters $r^{(1,2)} = N R^{(1,2)}/\left( 1-R^{(1,2)}\right)$, 
the  upper bound \eqref{Fmax} changes to $ \mathcal{F} \leqslant  4 N^2 / ( \sqrt{r^{(1)}} + \sqrt{r^{(2)}} )^2 = \mathcal{F}_{\textrm{upper}} $.

The optimal wavefunction is computed by minimizing the correction to the Fisher
information [$x = n / N$, $x_\ast=\sqrt{r^{(1)}} / (\sqrt{r^{(1)}}+\sqrt{r^{(2)}})$]:
\begin{equation}
  \Delta \mathcal{F} \approx N^2 \int_0^1 \left[ \frac{(x -
  x_{\ast})^2}{\sqrt{r^{(1)} r^{(2)}}} \tilde\phi^2 (x) + \frac{4 \tilde\phi'^2 (x)}{\left(
  \sqrt{r^{(1)}} + \sqrt{r^{(2)}} \right)^4} \right] \dd x,
\end{equation}
which produces a Gaussian centered at $x=x_{\ast}$ of width $\sqrt{2} (r^{(1)} r^{(2)})^{1/8}/(\sqrt{r^{(1)}}+\sqrt{r^{(2)}})$.
This width scales as $N^{3/4}$ (cf. $N^{2/3}$ for single mode losses). 

\begin{figure}[!ht]
\includegraphics[width=3in]{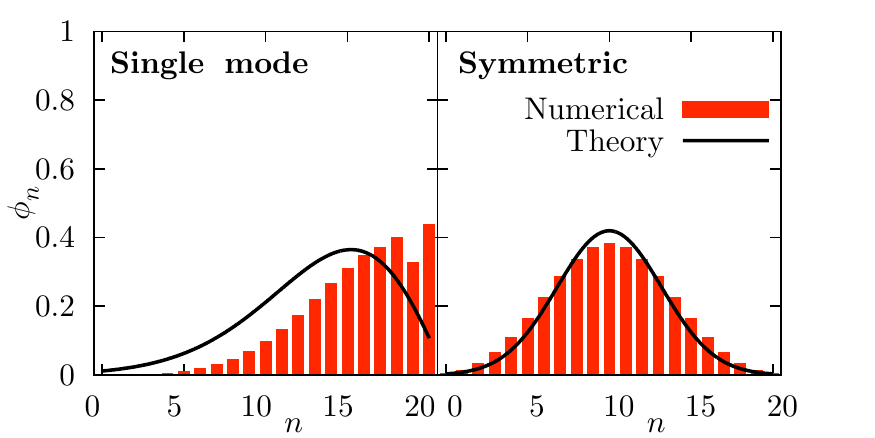}
\caption{\label{fig:phi}
Optimal 20-photon states for 95\% loss ($r=380$) in one (\emph{left}) and two (\emph{right}) modes.
Red bars represent amplitudes $\phi_n$ obtained by numerical optimization.
Black lines represent analytical approximation with an Airy function and a Gaussian.
These optimal states offer a precision improvement (square root of Fisher information) over coherent light 
of just 6\% (single-mode losses) and 0.4\% (symmetric losses), owing to high loss amount. 
(For single-mode losses, using coherent light, the precision used in calculation is for the
optimal reflectivity of the beamsplitter in MZ interferometer.)
}
\end{figure}

\begin{figure}[!ht]
\includegraphics[width=3in]{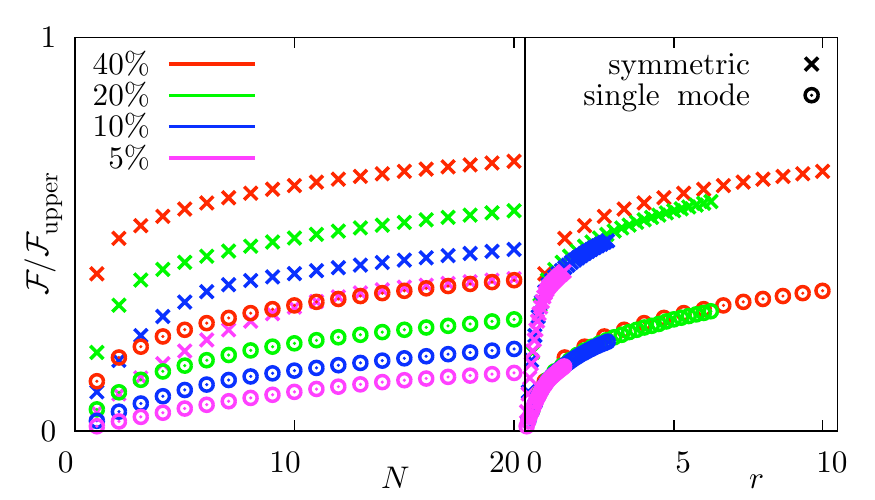}
\caption{\label{fig:cmp}
\emph{Left:} Fisher information for symmetric ($R^{(1)}=R^{(2)}=R$)
and single-mode ($R^{(1)}=R$, $R^{(2)}=0$) losses for optimal $N$-photon states
as a fraction of the linear upper bound [$N(1-R)/R$ and $4N(1-R)/R$ respectively].
The curves must tend to $1$ for large $N$, but the convergence is faster for losses in both arms.
For symmetric losses we use exact Fisher information, not the approximate upper bound.
\emph{Right:} Collapse of data when replotted as a function of $r=N R/(1-R)$ [see Eq.~(\ref{scaling})]}
\end{figure}

For moderate losses, this optimal form is attained when the number of photons is large.
In the limit of large losses, this asymptotic form is reached with a small 
number of photons (see Fig.~\ref{fig:phi}), within reach of current laboratory capabilities.
The Fisher information together with the leading correction is
\begin{equation}
  \mathcal{F}_{\textrm{opt}} = \frac{4 N^2}{\left( \sqrt{r^{(1)}} + \sqrt{r^{(2)}} \right)^2}
  \left[ 1 - \frac{2}{(r^{(1)} r^{(2)})^{1/4}} + O\left( \frac{1}{r}
    \right) \right] .
\end{equation}
The correction scales as $N^{1/2}$, in contrast to
the $N^{2/3}$ scaling for single mode losses. Correspondingly,
the crossover to the limiting behavior is expected for smaller $N$.
The convergence to asymptotic precision for the case of single mode and 
symmetric losses is illustrated in Fig.~\ref{fig:cmp}

\emph{Summary and Outlook:} An important question in interfer-
ometry is that of scaling of precision with photon number N,
and whether this scaling changes as N increases. We have shown analytically that the best supra-classical precision, as quantified by the QFI, 
is quadratic in $N$ initially and undergoes the crossover into the regime where it 
scales linearly with $N$. This has far-reaching repercussions --- the crossover location $N_c$ is a function only of dissipation, 
due to these losses a resource bound (or scale) is imposed on any
instrument claiming to offer supra-classical precision. Our results, indicating that
the improvement of precision is dissipation-limited, should temper expectations of various proposals advertising a Òquantum leapÓ in sensitivity of gravita-
tional wave detectors with non-classical light \cite{Caves81,review}.

We obtained the scaling relationship for the Fisher information and
studied the evolution of the optimal state: it undergoes a sequence
of bifurcations for intermediate values of $N$. In the limit of very
large $N$, the optimal state becomes a continuous function with width
that scales as $N^{2/3}$ or $N^{3/4}$ (for one/two mode loss), a signature
property that makes it distinct from states considered previously.

It should be noted that the precise value of the exponent $\epsilon$ for the width scaling (or the exact form
of the wavefunction) does not affect the asymptotic value of the Fisher information. It is vital, however, that the width be much greater than $\sqrt{N}$
and much smaller than $N$. Any exponents lying strictly within $1/2 < \epsilon <1$ will give the asymptotic 
value of the Fisher information, though convergence rates will be optimized with $2/3$ or $3/4$. The spin coherent state discussed in 
Ref.~\cite{DowlingNumerical} with respect to the large
loss limit has a width scaling $\propto N^{1/2}$ and is
suboptimal -- it does not offer supra-classical precision for any losses
in view of Caves' theorem \cite{Caves81}. Similarly, the Holland-Burnett state \cite{Holland93} with width $\propto N$ 
is also suboptimal: for symmetric losses its asymptotic Fisher information is half the upper bound.

As a final note, we enlarge upon two proposed applications of quantum
light; to free-space target acquisition and ranging, and to gravity wave observation. In clear weather,
infra-red light is attenuated by $0.5 - 1.0$dB/km.
Locating a target at $10$km distance requires $20$km of
roundtrip propagation, i.e. $R= 90 - 99\%$ loss,  combined with a typical $10\%$ target 
reflectance becomes $99 - 99.9\%$. Optimal Fisher information per \emph{received} photon is $4/R$, 
thus naively one would expect a two-fold improvement in phase precision $\delta \phi$ over 
coherent light (having Fisher Information unity per received photon) even for high losses.
This comparison is for an interferometer with 50/50 beamsplitters;
precision can be trivially increased with coherent light inputs by optimizing the 
beamsplitter reflectances. Compared with this strategy, non-classical light 
can improve precision by at most a factor of $(1+\sqrt{1-R})/\sqrt{R}$, i.e. by $3 - 10\%$ for losses above.
This fractional advantage in the very high loss limit does not offset the high practical cost of generating 
those optimal states we have discovered. This result should moderate expected outcomes of such proposals. 

To contrast, consider two-mode losses $R \approx 1\%$, the expected domain of advanced 
interferometric gravitational wave detectors with high-reflectivity mirrors and state-of-the-art photodetectors.
The improvement to  $\delta \phi$ over classical light for the same $N \gg 1$ approaches
a factor of $1/\sqrt{R}$. This 10-fold improvement falls far short of more optimistic estimates
assuming idealized conditions \cite{review} but still represents a clear, 
non-trivial advantage for the optimal input states we have discovered. 
The ability to reduce intrinsic quantum noise by an order of magnitude without an associated 
increase in radiation pressure noise (the photon flux has not increased) is certainly of interest for the development of gravity wave detectors.

We remark that due to the isomorphisms between two harmonic oscillators and spin/qubit systems these results apply 
quite broadly in metrology protocols, from photonic systems and atomic condensates, to spin ensembles coupled to heat 
baths, and other processes undergoing both unitary evolution and dissipation.

G.A.D. contributed to this work while under contract with Mission Critical Technologies, Inc.


\begin{thebibliography}{99}
\bibitem{footnote2}
An additional contribution from the boundary is $\propto \phi^2(0) / r^{7/6}$,
which would be much larger than $\Delta \mathcal{F}$ unless $\phi(0)=0$.
\end{thebibliography}
\end{document}